\documentclass{article}
\usepackage{spconf,amsmath,amssymb,graphicx,xcolor,booktabs,url}

\newcommand{\method}{MuSE}   % method name -- change here
\newcommand{\cA}{\mathcal{A}}
\newcommand{\cK}{\mathcal{K}}
\newcommand{\cP}{\mathcal{P}}
\newcommand{\cS}{\mathcal{S}}

\newcommand{\bx}{\mathbf{x}}
\newcommand{\bv}{\mathbf{v}}
\newcommand{\bb}{\mathbf{b}}
\newcommand{\bu}{\mathbf{u}}
\newcommand{\bz}{\mathbf{z}}
\newcommand{\bn}{\mathbf{n}}
\newcommand{\bU}{\mathbf{U}}
\newcommand{\bV}{\mathbf{V}}
\newcommand{\bX}{\mathbf{X}}

\title{High-Resolution Quantitative MRI Using\\ Contrast-Domain Multiscale Energy (M\lowercase{u}SE) Regularization}

\name{Rushdi Rusho$^{1}$, Jyothi Rikhab Chand$^{1}$, Steve R. Kecskemeti$^{2}$, James H. Holmes$^{3}$, Mathews Jacob$^{1}$}
\address{$^{1}$University of Virginia, Charlottesville, VA, USA\\
$^{2}$University of Wisconsin--Madison, Madison, WI, USA\\
$^{3}$University of Iowa, Iowa City, IA, USA}

\begin{document}
\maketitle

% =========================================================================
\begin{abstract}
Quantitative MRI methods can offer tissue parameter maps (e.g. $T_1$) that are comparable across scanners, sessions and subjects, together with multiple contrast-weighted images from a single scan. However, most data are acquired at 1\,mm, which limits their use in neurodegenerative diseases that affect fine brain structures. We introduce, to our knowledge, the first diffusion-model-based reconstruction for quantitative MRI at 0.5\,mm isotropic resolution from scans of less than 5 minutes, which has eightfold smaller voxel volume than clinical protocols. The key idea is to apply the prior to the synthesized contrast images rather than to the subspace coefficients. The prior is then independent of the temporal basis and the sequence, which improves generalization and allows a 2-D multiscale energy (\method) model trained on grayscale CelebA face images to be used without any MRI training data. The explicit energy makes reconstruction MAP estimation by a step-size-free majorize--minimize algorithm that regularizes a rotating subset of contrasts along different axes, which makes 3-D multi-contrast reconstruction tractable. For 3-D radial MPnRAGE at 7\,T in 4.3\,min, \method\ reduced the phantom NRMSE from 0.133 (CG-SENSE) and 0.091 (L1-wavelet) to 0.074 and the WM $T_1$ interquartile range from 117 and 47\,ms to 38\,ms, and it was demonstrated in vivo at 7\,T.
\end{abstract}


% =========================================================================
\section{Introduction}
\label{sec:intro}

Quantitative MRI estimates tissue parameters such as $T_1$ that are more readily comparable across scanners, sessions and subjects. Methods such as MPnRAGE fit a signal model to images acquired during magnetization recovery \cite{kecskemeti2016mpnrage}. In neurodegenerative diseases, early changes affect thin structures, such as the cortical ribbon and the hippocampal subfields, whose dimensions approach 1\,mm \cite{wisse2021hippocampal}. However, a 0.5\,mm isotropic voxel has one eighth of the volume of a 1\,mm voxel; because the signal-to-noise ratio scales with voxel volume and the square root of scan time, preserving it would require 64 times the scan time.

Subspace models are the most widely used approach to accelerate quantitative MRI \cite{liang2007spatiotemporal}. They constrain the time profile of each voxel to a few principal components of Bloch-simulated signal evolutions, which leaves a few spatial factors to be estimated. These spatial factors can be further regularized by wavelet sparsity \cite{lustig2007sparse} or locally low-rank (LLR) penalties \cite{zhang2015accelerating}. These methods degrade at high acceleration and low SNR (Sec.~\ref{sec:reg}). Diffusion models, which learn multiscale denoisers from images, are now emerging as powerful priors for MRI reconstruction \cite{chung2022score}. Most perform posterior sampling, which typically needs many network evaluations; extending them to 3-D quantitative imaging with hundreds of contrast volumes is challenging. In addition, fully sampled training data are impractical to acquire at this resolution.

We introduce, to our knowledge, the first diffusion-model-based reconstruction for quantitative MRI at 0.5\,mm resolution. It rests on three design choices. First, we use multiscale energy (\method) models \cite{chand2024multi,chand2026annealed}, diffusion models whose denoiser is the gradient of an explicit energy. The explicit energy poses reconstruction as maximum a posteriori (MAP) estimation, which we solve by majorize--minimize (MM) with quadratic, step-size-free updates. This avoids the posterior sampling used by approaches such as DPS and DAPS \cite{chung2023diffusion,zhang2025daps}. Second, to minimize the risk of overfitting limited, basis-specific training data, we apply the prior to the contrast images synthesized from the subspace model \cite{chen2025ssmuse} rather than to the spatial factors. The prior is then independent of the basis and the sequence, so a generic 2-D model learned from grayscale face images (CelebA) can be used without MRI training data; such generic priors were recently shown to generalize well across imaging domains \cite{glaszner2025bigger}. Third, we apply this 2-D prior to slices along all three orientations \cite{lee2023improving}, and each stochastic MM iteration regularizes a rotating subset of contrasts along different axes, which makes hundreds of 3-D contrasts tractable.

We evaluate \method\ on 3-D radial MPnRAGE data at 7\,T (8.6\,min, and a $2\times$ subset with 4.3\,min) against CG-SENSE, L1-wavelet and LLR, on a phantom and in vivo.

\begin{figure}[t]
    \centering
    \includegraphics[width=0.97\columnwidth]{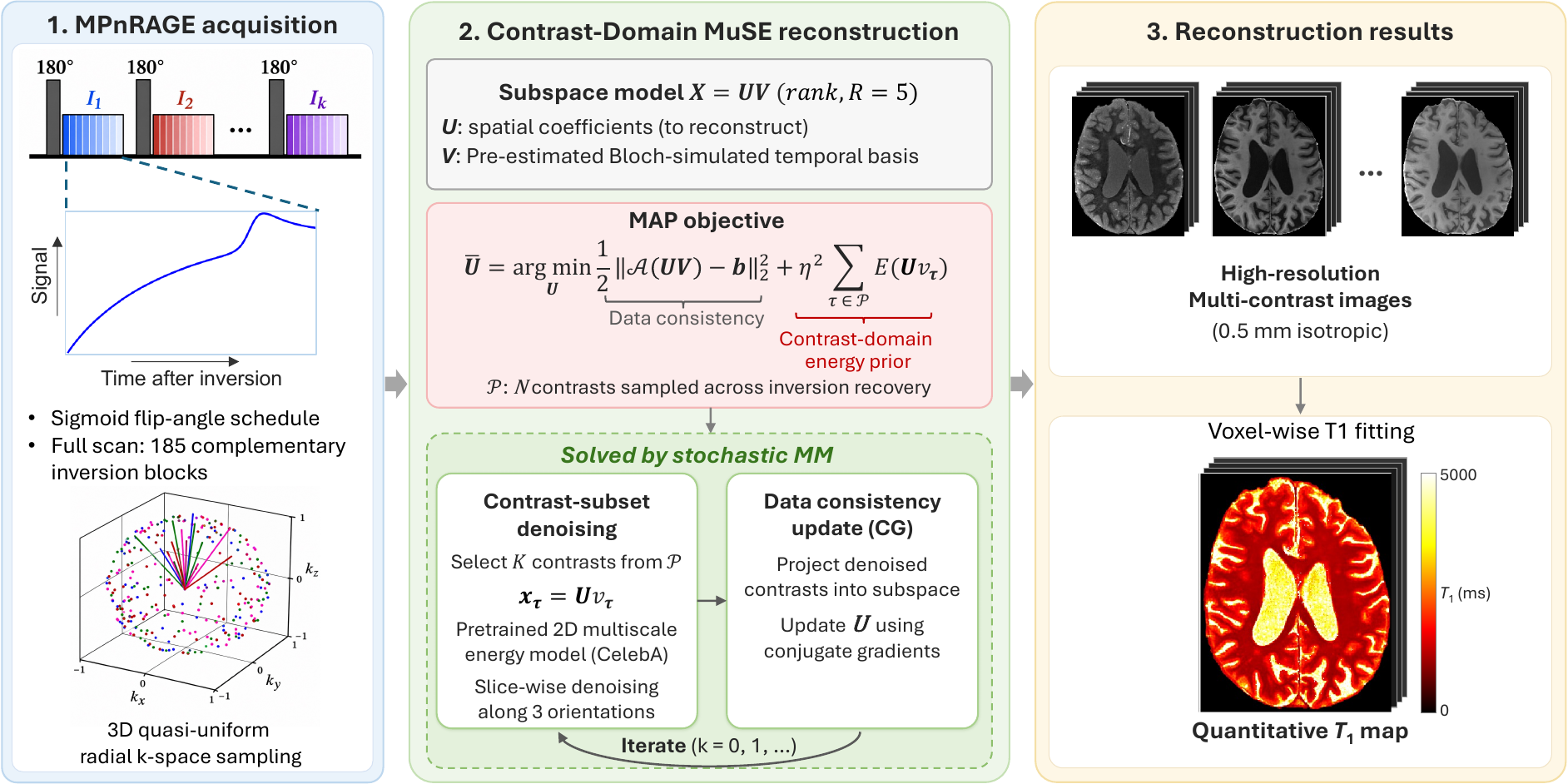}
    \caption{Overview of \method\ for 0.5\,mm quantitative MRI. (1)~3-D radial MPnRAGE acquisition. (2)~Each MM iteration denoises a subset of synthesized contrasts slice by slice with a 2-D \method\ model trained only on grayscale CelebA faces, then enforces data consistency by CG in the rank-five subspace. (3)~Voxel-wise $T_1$ fitting at 0.5\,mm.}
    \label{fig:overview}
\end{figure}

% =========================================================================
\section{Background}
\label{sec:background}

\subsection{MPnRAGE acquisition}
We use an MPnRAGE \cite{kecskemeti2016mpnrage} sequence,  on a 7\,T scanner, which repeats an inversion block (Fig.~\ref{fig:overview}, step~1). Each block consists of an inversion pulse, 461 3-D radial spokes (TR 4.94\,ms, 2.3\,s; the first is discarded) and a 499-ms recovery delay, about 2.8\,s in total. The flip angle rises from $4^\circ$ to $8^\circ$ along a sigmoid. Spokes with the same index in different blocks share the same time since inversion, so each of the $T=460$ spoke indices defines one contrast $\tau$, sampled by the spokes of all blocks. The 185 blocks take 8.6\,min; the last 92 blocks give the $2\times$ scan (4.3\,min).

\subsection{Subspace model} The contrast images $\bX=[\bx_1,\dots,\bx_T]\in\mathbb{C}^{M\times T}$ follow $\bx_\tau(m)=\rho_m\,s_\tau(\vartheta_m)$, where $\rho_m$ is the proton density and $s_\tau(\vartheta_m)$ the magnetization at time $\tau$. In this work, $\vartheta_m$ is a function of $T_1$, $B_1^{+}$ and the inversion efficiency (IE) \cite{kecskemeti2016mpnrage}. The measurements are $\bb_\tau=\cA_\tau\bx_\tau+\bn_\tau$, where $\cA_\tau$ includes the coil sensitivities, the trajectory and rigid motion, and $\bn_\tau$ is Gaussian noise of variance $\nu^2$. Subspace methods write $\bX=\bU\bV$ \cite{liang2007spatiotemporal}, where $\bV\in\mathbb{R}^{R\times T}$ has orthonormal rows estimated from Bloch simulations and $\bU\in\mathbb{C}^{M\times R}$ holds the spatial coefficients, so that $\bx_\tau=\bU\bv_\tau$.

\subsection{Classical regularizers}
\label{sec:reg}
L1-wavelet regularization penalizes the $\ell_1$ norm of the wavelet detail coefficients of the columns of $\bU$ \cite{lustig2007sparse}. LLR soft-thresholds the singular values of the $P\times R$ matrix of coefficients in each block of $P$ voxels \cite{zhang2015accelerating}. If the block has local rank $k\le R$, the rank-$k$ approximation retains a fraction of about $k/R$ of the noise energy for $P\gg R$, so with $R=5$ the denoising is limited unless $k$ is well below $R$.

\subsection{Multiscale energy models}
The prior of 2-D images $\bu$ at noise level $\sigma$ is $p_w(\bu;\sigma)\propto\exp(-\phi_w(\bu;\sigma)/\sigma^2)$, with $\phi_w(\bu;\sigma)=\frac12\|\bu-\psi_w(\bu;\sigma)\|_2^2$ and a convolutional network $\psi_w$ \cite{chand2024multi,chand2026annealed}. Its gradient defines the denoiser
\begin{equation}
D_\sigma(\bu) = \bu - \nabla_{\bu}\phi_w(\bu;\sigma),
\label{eq:tweedie}
\end{equation}
trained by multiscale denoising score matching \cite{vincent2011connection}, which makes $D_\sigma$ approximate the MMSE denoiser. The explicit energy allows classical optimization, such as the majorize--minimize scheme below, instead of posterior sampling. Annealing $\sigma$ from large to small further improves convergence: early iterations smooth the energy and fix coarse structure, and later ones recover detail \cite{chand2024multi,chand2026annealed}.

% =========================================================================
\section{Proposed Method}
\label{sec:methods}

\subsection{Multiscale MAP formulation using contrast priors}
\textbf{Contrast prior.} Unlike SS-MuSE \cite{chen2025ssmuse}, which learns a prior directly on the spatial factors $\bU$, we learn it indirectly through the contrast images $\bx_\tau=\bU\bv_\tau$, $\tau=1,\dots,T$, which are linear functions of $\bU$. A prior on $\bU$ requires training data specific to the basis functions $\bV$, which unfortunately is not available in our high-resolution setting. A prior on contrast images can instead be learned from generic images, and such priors were recently shown to generalize well to MRI \cite{glaszner2025bigger}.

\textbf{Slice-wise energy.} A 3-D energy would need much 3-D training data and memory, so we apply the 2-D model to slices. For a volume $\bx\in\mathbb{C}^{n\times n\times n}$ and an axis $a\in\{x,y,z\}$, the operator $\cS_a$ extracts the $n$ 2-D slices perpendicular to $a$, each an $n\times n$ image, and the real and imaginary parts are treated as separate images \cite{lee2023improving,chen2025ssmuse}. The energy of the volume along $a$ is the sum of the energies of its slices, $e_\sigma(\cS_a\bx)=\sum_{j}e_\sigma(\bx_{a,j})$, where $e_\sigma=-\log p_w(\cdot;\sigma)$ and $\bx_{a,j}$ is the $j$th slice.

\textbf{MAP estimation.} We estimate $\bU$ by minimizing
\begin{equation}
\begin{aligned}
J(\bU) &= \tfrac12\|\cA(\bU\bV)-\bb\|_2^2 + \eta^2 E_\sigma(\bU),\\
E_\sigma(\bU) &= \sum_{\tau\in\cP}\sum_{a\in\{x,y,z\}} e_\sigma\big(\cS_a(\bU\bv_\tau)\big),
\end{aligned}
\label{eq:map}
\end{equation}
where $\eta$ is a regularization parameter and $\cP$ is a pool of $N$ frames spread over the inversion recovery. Since $R\ll T$, a pool much smaller than $T$ suffices. By \eqref{eq:tweedie}, $\nabla e_\sigma(\bx)=(\bx-D_\sigma(\bx))/\sigma^2$, so $E_\sigma$ is never evaluated explicitly.

\subsection{Majorization} For an ideal MMSE denoiser, the Hessian of $e_\sigma$ is $\sigma^{-2}\big(I-\sigma^{-2}\mathrm{Cov}[\bu\,|\,\tilde\bu]\big)\preceq\sigma^{-2}I$ \cite{manor2024posterior}, so at $\bx_\tau^k=\bU_k\bv_\tau$ each term is bounded by a quadratic majorizer whose minimization requires no step size (for the trained network, this bound holds approximately),
\begin{equation}
e_\sigma(\bx) \le \tfrac{1}{2\sigma^2}\big\|\bx - D_\sigma(\bx_\tau^k)\big\|_2^2 + c_\tau^k.
\label{eq:major}
\end{equation}

\subsection{Rotating contrast subsets}
Evaluating all $3N$ terms at every iteration is expensive, so at iteration $k$ we keep the terms of a subset $\cK_k\subseteq\cP$ of $K\le N$ frames, each assigned one axis $a_\tau$, and scale each kept term by $3N/K$ so that the sum approximates the full sum over $3N$ terms. The subset and the axis assignments rotate between iterations, so that \method\ is applied along different axes and, over the iterations, to all frames of the pool. The resulting majorizer is stochastic: it is an unbiased surrogate of \eqref{eq:major} summed over all terms if the frames and axes are drawn uniformly \cite{mairal2013stochastic}, and rotating through interleaved subsets approximates this. Each iteration costs $K$ 2-D passes over the volume, and the descent is not monotone. The selected contrasts are denoised, $\hat{\bx}_\tau=D_\sigma(\bx_\tau^k)$ along $a_\tau$, and mapped back to the subspace by least squares,
\begin{equation}
\tilde{\bU}_k = \bU_k + \!\sum_{\tau\in\cK_k}\!\big(\hat{\bx}_\tau - \bx_\tau^k\big)\bv_\tau^{H}\big(\bV_{\cK_k}\bV_{\cK_k}^{H}+\epsilon I\big)^{-1},
\label{eq:lsq}
\end{equation}
where $\bV_{\cK_k}$ holds the columns of $\bV$ of the selected frames and $\epsilon>0$ is a small constant that stabilizes the inverse. Approximating $\|(\bU-\tilde{\bU}_k)\bV_{\cK_k}\|_F$ by $\|\bU-\tilde{\bU}_k\|_F$ and absorbing constant scales into $\eta$ gives
\begin{equation}
\bU_{k+1} = \arg\min_{\bU}\; \tfrac12\|\cA(\bU\bV)-\bb\|_2^2 + \tfrac{\lambda_k}{2}\|\bU-\tilde{\bU}_k\|_F^2,
\label{eq:dc}
\end{equation}
with $\lambda_k=\eta^2/\sigma_k^2$, solved by a few warm-started conjugate gradient (CG) iterations.

\subsection{Continuation and renoising}
As $\sigma_k$ is annealed, $\lambda_k$ increases, so the weight on mildly denoised anchors grows. The denoiser is trained on inputs of the form $\bu+\sigma\bz$, whereas the iterate $\bx_\tau^k$ carries residual aliasing and noise that are not Gaussian at level $\sigma_k$. To reduce this train--test mismatch, as in D-MOD \cite{ahmed2026dmod}, the anchor can be averaged over $L$ renoised inputs, $\hat\bx_\tau=\frac1L\sum_{l}D_\sigma(\bx_\tau^k+\sigma\bz_l)$ with $\bz_l\sim\mathcal{N}(0,I)$; $L=0$ means no renoising.

% =========================================================================
\section{Experiments}
\label{sec:impl}

\subsection{Validation strategy and BrainWeb phantom}
No ground truth is available for 0.5\,mm quantitative MRI in vivo, so we validated the methods on a numerical phantom with known $T_1$ maps and contrasts. The BrainWeb anatomical model \cite{cocosco1997brainweb} provides 1\,mm segmentation maps, which we interpolated to 0.5\,mm. We assigned representative $T_1$, $B_1^{+}$, inversion efficiency (IE) and proton density (PD) values to gray matter, white matter, CSF and skull, which gives the parameter maps $\vartheta_m$ and $\rho_m$ of the signal model in Sec.~\ref{sec:background}. The ground-truth contrasts ($T=460$) were generated by Bloch simulation of the MPnRAGE sequence with the blocks, spoke timing and flip-angle schedule of the in vivo scan. The k-space data were then simulated with the forward model $\cA_\tau$, that is, the in vivo radial trajectory and coil sensitivities, without motion, with in vivo-matched noise added.

\subsection{In vivo data}
One healthy volunteer was scanned on a 7\,T GE scanner with a 32-channel head coil (Sec.~\ref{sec:background}). Images were reconstructed at 0.5\,mm isotropic resolution ($512^3$, 256\,mm FOV) from 16 compressed virtual coils, with per-block rigid motion included in $\cA_\tau$.

\subsection{Training and use of the prior}
The network $\psi_w$ is a 3.4\,M-parameter SongUNet with EDM preconditioning \cite{karras2022elucidating}. It was trained by denoising score matching on 100\,k grayscale CelebA face images, with no MRI data. To apply it to MRI, each synthesized contrast volume is cut into 2-D slices along the chosen axis. Because the network is trained on real-valued images, the real and imaginary parts of each slice are denoised separately as two grayscale images and then recombined into the complex slice, $\hat{\bx}=D_\sigma(\mathrm{Re}\,\bx)+i\,D_\sigma(\mathrm{Im}\,\bx)$.

\subsection{Compared methods, tuning and metrics}
All methods share the basis ($R=5$), the forward model (scaled to $\lambda_{\max}(\cA^H\cA)=1$) and the coil maps. CG-SENSE \cite{pruessmann2001advances} uses 50 CG iterations and initializes the regularized methods. \method\ uses a pool of $N=8$ frames (frames 10--420, TI 69--2095\,ms) with \method\ applied along rotating axes (here $K=N$, so the subset is the full pool), no renoising, and the first seven levels of a ten-level EDM schedule from $\sigma_0$ (the noise level of the CG-SENSE images) to $0.08\,\sigma_0$. Each \method\ iteration uses 20 CG iterations. L1-wavelet uses a six-level orthonormal Daubechies-4 transform and ADMM (10 iterations of 5 CG steps); LLR uses shifted $8^3$ blocks, the threshold $c\,(\sqrt{P}+\sqrt{R})$ in noise units and ADMM (10 iterations). At $1\times$ and $2\times$, the \method\ weight is $\eta/\sigma_0=0.2$ and $0.15$, the wavelet weight is $\lambda=3.2\times10^{-3}$ and $1.6\times10^{-3}$, and the LLR constant is $c=0.25$ and $0.15$.

The parameters of all methods were tuned on the phantom to minimize the NRMSE in the brain, averaged over the WM-nulled, MPRAGE-like and CSF-nulled frames (TI 415/1398/855\,ms). In vivo, the noise-relative $c$ and $\eta$ were kept, and the wavelet $\lambda$ was scaled by the ratio of noise levels and data gains (0.69 at $1\times$, 0.82 at $2\times$). Figures show contrasts synthesized with $B_1^{+}=\mathrm{IE}=1$. Fine detail was assessed by the high-frequency error norm (HFEN; 3-D LoG, $\sigma=1.5$ voxels) in the brain, and $T_1$ by the median bias and interquartile range (IQR) in WM and GM on three orthogonal slices.

% =========================================================================
\section{Results and Discussion}
\label{sec:results}

\begin{table}[t]
\centering
\caption{Phantom accuracy (lower is better). NRMSE and high-frequency error (HFEN) over the brain volume, averaged over the WM-nulled, MPRAGE-like and CSF-nulled contrasts, at $1\times$ (8.6\,min) and $2\times$ (4.3\,min). $T_1$ rows: median bias/IQR (ms) at $2\times$. \method\ has the lowest NRMSE and HFEN in every row, and its $2\times$ NRMSE is below the $1\times$ L1-wavelet value.}
\label{tab:phantom}
\footnotesize
\setlength{\tabcolsep}{4pt}
\begin{tabular}{@{}lcccc@{}}
\toprule
 & CG-SENSE & LLR & L1-wavelet & \method \\
\midrule
NRMSE, $1\times$ & 0.119 & 0.105 & 0.080 & \textbf{0.058} \\
HFEN, $1\times$  & 0.197 & 0.217 & 0.253 & \textbf{0.146} \\
NRMSE, $2\times$ & 0.133 & 0.110 & 0.091 & \textbf{0.074} \\
HFEN, $2\times$  & 0.317 & 0.249 & 0.274 & \textbf{0.244} \\
$T_1$ bias/IQR, WM & -42/117 & -6/66 & \textbf{-2}/47 & -19/\textbf{38} \\
$T_1$ bias/IQR, GM & -34/325 & -34/251 & -30/193 & \textbf{-18}/\textbf{159} \\
\bottomrule
\end{tabular}
\end{table}

\begin{figure}[t]
    \centering
    \includegraphics[width=0.93\columnwidth]{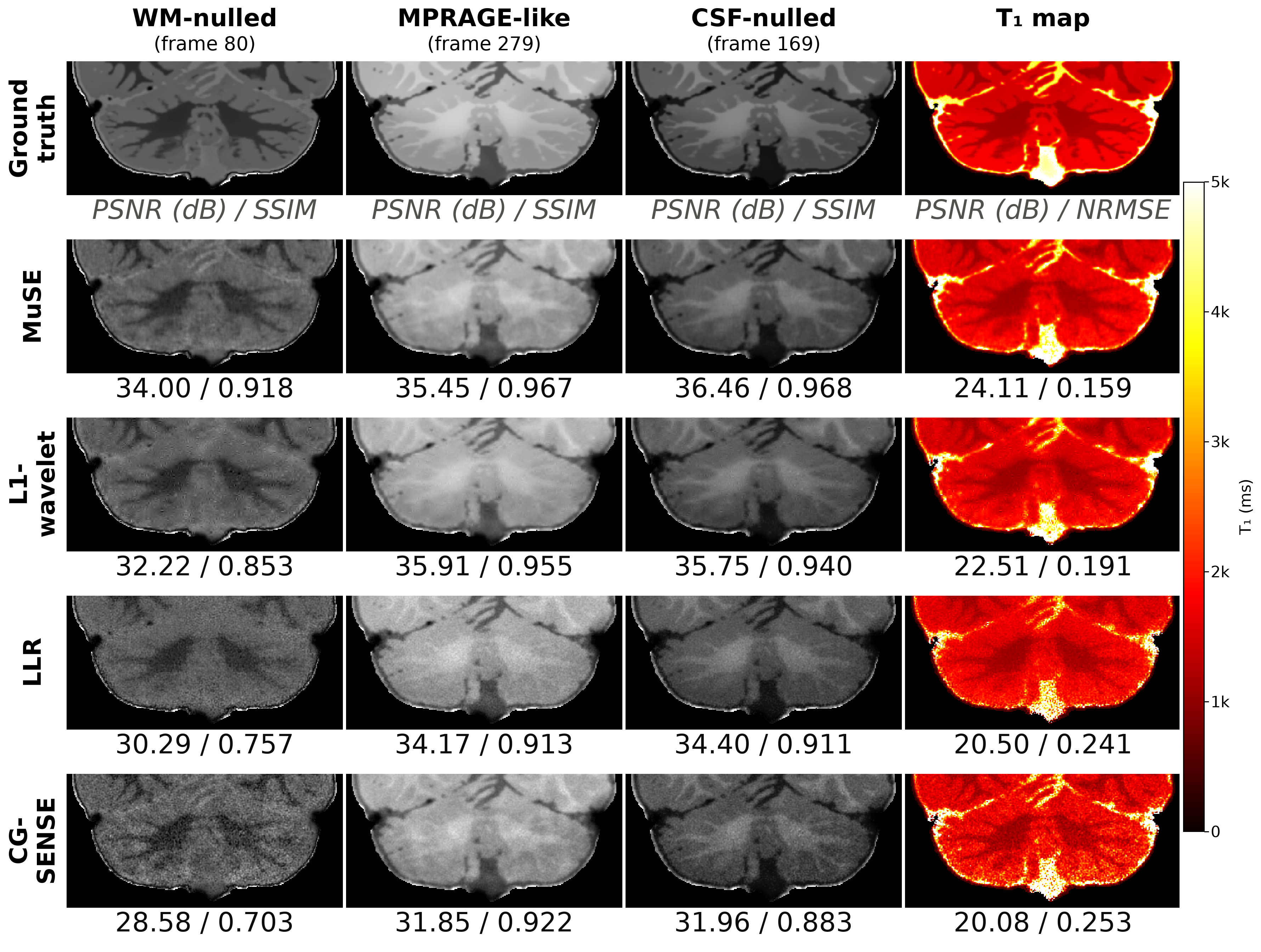}
    \caption{BrainWeb validation at $2\times$ (4.3\,min), cerebellar crop, compared with the ground truth. Columns: three synthesized contrasts (frames 80, 279, 169) and the $T_1$ map; metrics under each panel are PSNR/SSIM (contrasts) and PSNR/NRMSE ($T_1$) over the full-slice brain mask.}
    \label{fig:phantom_fig}
\end{figure}

\begin{figure}[!tb]
    \centering
    \includegraphics[width=1.0\columnwidth]{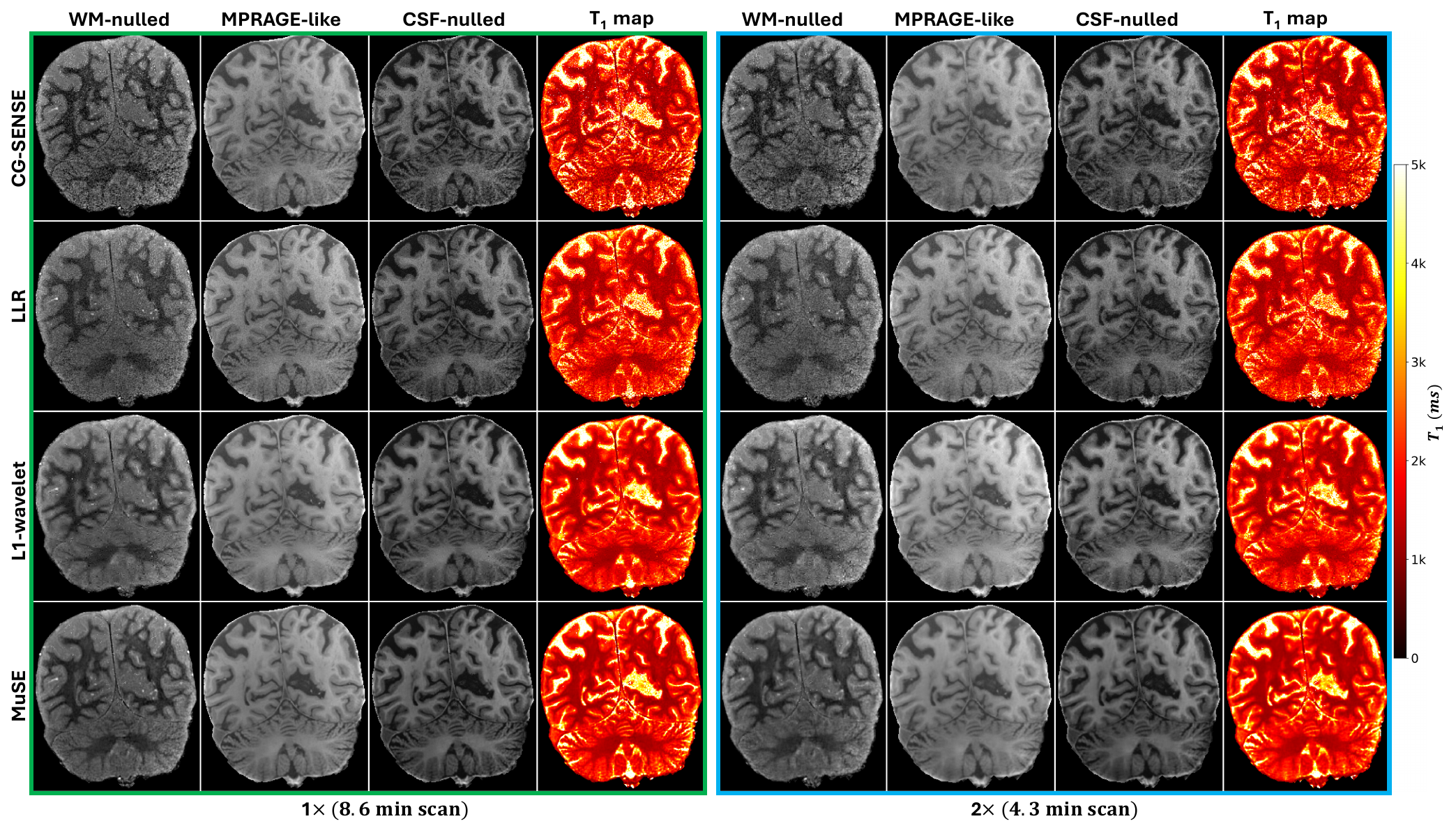}
    \caption{In vivo reconstructions at $1\times$ (8.6\,min) and $2\times$ (4.3\,min), layout as in Fig.~\ref{fig:phantom_fig}; no ground truth is available.}
    \label{fig:invivo_comparison}
\end{figure}

\textbf{Phantom comparison.} \method\ gave the lowest whole-brain NRMSE at both scan times (Table~\ref{tab:phantom}); at $2\times$, it was 44\% lower than CG-SENSE and 19\% lower than L1-wavelet, and even below the $1\times$ L1-wavelet result (0.074 versus 0.080). It also had the lowest HFEN and, at $2\times$, the narrowest $T_1$ IQR in WM (38\,ms, versus 47--117) and GM (159\,ms, versus 193--325). L1-wavelet was the closest competitor, and the two methods are nearly equal on the high-SNR MPRAGE-like frame (NRMSE 0.0426 versus 0.0431; PSNR 35.9 versus 35.5\,dB). The difference lies in the low-SNR images (Fig.~\ref{fig:phantom_fig}). In the WM-nulled frame, L1-wavelet leaves a mottled, patchy texture, whereas \method\ is smoother and keeps the thin cerebellar white matter branches (PSNR 34.0 versus 32.2\,dB; SSIM 0.918 versus 0.853). The $T_1$ map shows the same pattern: L1-wavelet is speckled, whereas \method\ is homogeneous with sharper folia boundaries (NRMSE 0.159 versus 0.191). \method\ had a WM $T_1$ bias of $-19$\,ms at $2\times$.

\textbf{In vivo.} Fig.~\ref{fig:invivo_comparison} shows the in vivo reconstructions of the four methods at $1\times$ and $2\times$. Since no ground truth exists, we compare them only visually, and a quantitative in vivo evaluation is left for future work.

\section{Limitations}
This study is preliminary: it uses one subject and one phantom, the in vivo comparison is visual, memory use is high (about 115\,GB on each of two GPUs at $1\times$), and the face-trained prior is untested in pathology. More elaborate studies will follow.

\section{Conclusion}
\label{sec:conclusion}

\method\ regularizes subspace quantitative MRI with a multiscale energy prior on synthesized contrasts, learned from face images and minimized by a step-size-free stochastic MM algorithm. At 0.5\,mm, it outperformed the other methods on a phantom and was demonstrated in vivo at 7\,T.

% =========================================================================
\section{Compliance with Ethical Standards}
This study was performed in line with the principles of the Declaration of Helsinki. In vivo data were acquired at the University of Iowa with approval of the Institutional Review Board and written informed consent from the subject.

\section{Acknowledgments}
Supported by NIH grants R01-AG067078, R01-EB031169 and R01-EB019961. We thank UVA Research Computing (\url{rc.virginia.edu}) for computational resources. Claude (Anthropic) was used to refine the text of this paper. The authors reviewed the content and take full responsibility for it.

% =========================================================================
\clearpage
\bibliographystyle{IEEEbib}
\bibliography{isbi2027}

\end{document}